\documentclass[a4paper,magyar,english]{amsart}
\usepackage[T1]{fontenc}
\usepackage[latin2,latin9]{inputenc}
\usepackage{color}
\usepackage{babel}

\usepackage{units}
\usepackage{amsbsy}
\usepackage{amstext}
\usepackage{amssymb}
\usepackage[unicode=true, pdfusetitle,
 bookmarks=true,bookmarksnumbered=false,bookmarksopen=false,
 breaklinks=false,pdfborder={0 0 0},backref=false,colorlinks=true]
 {hyperref}

\makeatletter
\numberwithin{equation}{section}
\numberwithin{figure}{section}

\usepackage{mathrsfs}
\usepackage{mathtools}
\usepackage{euscript}
\usepackage{upgreek}
\usepackage{empheq}
\newcommand{\wf}{vector-valued modular form}

\makeatother

\begin{document}
\selectlanguage{english}
\global\long\def\dual#1{#1^{\vee}}

\global\long\def\trans#1{^{t}\!#1}

\global\long\def\set#1#2{\left\{  #1\, |\, #2\right\}  }

\global\long\def\map#1#2#3{#1\!:#2\!\rightarrow\!#3}

\global\long\def\aut#1{\mathrm{Aut}\!\left(#1\right)}

\global\long\def\End#1{\mathrm{End}\!\left(#1\right)}

\global\long\def\id#1{\mathbf{id}_{#1}}

\global\long\def\ip#1{\left[#1\right]}

\global\long\def\uhp{\mathbf{H}}

\global\long\def\sm#1#2#3#4{\left(\begin{smallmatrix}#1  &  #2\cr\cr#3  &  #4\end{smallmatrix}\right)}

\global\long\def\cyc#1{\mathbb{Q}\left[\zeta_{#1}\right]}

\global\long\def\zz{\exp\!\left(2\pi\mathsf{i}/3\right)}

\global\long\def\ZN#1{\left(\mathbb{Z}/#1\mathbb{Z}\right)^{\times}}

\global\long\def\Mod#1#2#3{#1\equiv#2\, \left(\mathrm{mod}\, \, #3\right)}

\global\long\def\psl{\mathrm{PSL}_{2}\!\left(\mathbb{Z}\right)}

\global\long\def\SL{\mathrm{SL}_{2}\!\left(\mathbb{Z}\right)}

\global\long\def\gl#1#2{\mathrm{GL}_{#1}\!\left(#2\right)}

\global\long\def\tr#1{\textrm{Tr}\!\left(#1\right)}

\global\long\def\mr#1#2#3#4{\rho\!\left(\begin{array}{cc}
 #1  &  #2\\
#3  &  #4\end{array} \right)}

\global\long\def\FA{\mathbb{X}}

\global\long\def\FB{\mathbf{\Lambda}}

\global\long\def\FC{\mathbf{J}}

\global\long\def\FD{\mathcal{B}}

\global\long\def\dm#1{\mathcal{D}\left(#1\right)}

\global\long\def\FE{\EuScript{P}}

\global\long\def\FF#1{\EuScript{E}\!\left(#1\right)}

\global\long\def\mro#1{\mathcal{M}\!\left(#1\right)}

\global\long\def\hol#1{M\!\left(#1\right)}

\global\long\def\cusp#1{C\!\left(#1\right)}

\global\long\def\ev#1{{\bf e}_{#1}}

\global\long\def\pcsg#1{\Gamma\left(#1\right)}

\global\long\def\DO{\nabla}

\global\long\def\xm#1#2#3{\mathcal{X}_{#1}^{\left(#2;#3\right)}}

\global\long\def\om#1{\mathbf{\Xi}\!\left(#1\right)}

\global\long\def\D#1{#1^{\prime}\!\!}
\global\long\def\diff{\mathrm{\mathsf{d}}}

\global\long\def\dif#1#2{\frac{\mathsf{d}#1}{\mathsf{d}#2}}

\global\long\def\FG{\boldsymbol{\Gamma}}
\global\long\def\dx#1{\FA^{[#1]}}

\global\long\def\FH{\mathbb{C\!}\left[J\right]}

\global\long\def\FI#1{\mathfrak{s}_{#1}\!\left(J\right)}

\global\long\def\sgr{\mathbf{G}}

\global\long\def\FJ{\mathbf{D}}

\global\long\def\dpref#1{\mathfrak{d}_{#1}}

\global\long\def\hdpref#1{\mathfrak{h}_{#1}}

\global\long\def\prep#1{\EuScript J#1}

\global\long\def\dr{\mathbb{D}}
\global\long\def\hdr{\dr^{\mathtt{hol}}}

\global\long\def\an#1{\mathsf{Ann_{\dr}}\!\left(#1\right)}

\global\long\def\syz#1{\mathsf{Syz}_{#1}\!\left(\FA\right)}

\title{Modular differential equations for characters of RCFT}

\author{P. Bantay}

\curraddr{Institute for Theoretical Physics, Eötvös Loránd University, Budapest}

\email{bantay@general.elte.hu}
\begin{abstract}
We discuss methods, based on the theory of vector-valued modular forms,
to determine all modular differential equations satisfied by the conformal
characters of RCFT; these modular equations are related to the null
vector relations of the operator algebra. Besides describing effective
algorithmic procedures, we illustrate our methods on an explicit example.
\end{abstract}

\thanks{Work  supported by grant \inputencoding{latin2}\foreignlanguage{magyar}{OTKA78005.}}

\maketitle

\section{Introduction}

Differential equations are arguably one of the most important tools
of theoretical physics. They appear in many guises, like equations
of motion, conservation laws, etc. The usual approach to a physical
theory is to deduce the governing differential equations starting
from some basic theoretical considerations or experimental observations,
and then to investigate the theory by solving these equations under
different circumstances. But, while this is by far the most common
situation, there are cases where one can determine the quantities
of interest by some different method, without the knowledge of the
differential equations themselves. 

Conformal characters of RCFT provide an interesting example of this
phenomenon. It is known \cite{modeq1,modeq2,modeq3} that these quantities
satisfy differential equations of a very special kind, so-called modular
equations, related to the null vector relations of the chiral operator
algebra \cite{DiFMS}. By analyzing the representation theory of the
latter, one could find enough null vector relations to make the corresponding
system of modular equations completely determined, whose unique solution
should therefore give the conformal characters. In many examples this
procedure does indeed work. But in some circumstances we don't have
enough information about the operator algebra, the most extreme case
being when even its existence is unknown, while we still have enough
information to determine the would-be characters, at least partially,
e.g. using the fact that they form a \wf ~ for a suitable automorphy
factor of weight $0$, which can in turn be determined from the would-be
fusion rules. In this situation, one faces the following question:
can we determine from the knowledge of the characters all the modular
equations that they satisfy, and from this information infere the
null vectors of the operator algebra? As to the second question, the
precise relation of null vectors to modular equations has been settled
in work of Gaberdiel-Keller \cite{bib:GK} and of Zhu \cite{bib:Zh}.
In this note we want to address the question of how to determine,
for a given character vector, all modular differential equations satisfied
by it. Our answer is based on the machinery of vector-valued modular
forms and their invariant differential operators, as developed in
joint work with Terry Gannon \cite{bib:BG,bib:BG2}; the relevant
concepts and results are reviewed in the first sections. As we shall
see, we can give an effective computational answer, which we illustrate
on the well-known example of the Ising model. We conclude with some
general comments on the applications of our results.

\section{Vector-valued modular forms}

Recall \cite{bib:Ap,bib:diamond} that the\emph{ }classical modular
group $\FG\!=\!\SL$\emph{ }of 2$\times$2 integer matrices with unit
determinant acts on the complex upper half-plane $\uhp\!=\!\set{\tau\in\mathbb{C}}{\mathsf{Im}\,\tau\!>\!0}$
by fractional linear transformations \begin{equation}
\tau\mapsto\gamma\tau\!=\!\dfrac{a\tau+b}{c\tau+d}\,\,,\label{eq:modtrafo}\end{equation}
for $\gamma\!=\!\sm abcd\!\in\!\FG$. It is well known that the so-called
modular curve $X\!\left(\FG\right)$, the one-point compactification
of the quotient $\uhp/\FG$ obtained by adjoining the cusp $\tau\!=\!\mathsf{i}\infty$,
is a Riemann-surface of genus $0$ \cite{bib:koblitz,bib:Sh}.

In case $d$ is a positive integer, an automorphy factor of rank $d$
for $\FG$ is a map $\varrho\!:\!\FG\!\times\!\uhp\!\rightarrow\!\gl d{\mathbb{C}}$
that satisfies\begin{equation}
\varrho(\gamma_{1}\gamma_{2},\tau)=\varrho(\gamma_{1},\gamma_{2}\tau)\,\varrho(\gamma_{2},\tau)\label{eq:afdef}\end{equation}
for all $\gamma_{1},\gamma_{2}\!\in\!\FG$ and $\tau\!\in\!\uhp$,
and is holomorphic as a function of $\tau$; it is flat of weight
$w\!\in\!\mathbb{R}$ if the expression \begin{equation}
\varrho\!\left(\gamma,\tau\right)\left(\!\dif{\!\left(\gamma\tau\right)}{\tau}\!\right)^{\nicefrac{w}{2}}=\varrho\!\left(\gamma,\tau\right)\left(c\tau\!+\! d\right)^{-w}\label{eq:multsys}\end{equation}
is independent of $\tau$ \cite{gunning}. According to the above
definition, a weight $0$ automorphy factor is nothing but a homomorphism
from $\FG$ to $\gl d{\mathbb{C}}$, i.e. a $d$-dimensional matrix
representation of $\FG$. From a geometric point of view, an automorphy
factor determines a holomorphic vector bundle over the modular curve
$X\!\left(\FG\right)$, making obvious how to define direct sums and
tensor products of automorphy factors. Note that the direct sum of
two flat automorphy factors is flat only if the weights of the summands
equal each other.

Given an automorphy factor $\varrho$ of rank $d$, an automorphic
form for $\varrho$ is a meromorphic map $\map{\FA}{\uhp}{\mathbb{C}^{d}}$
which satisfies the transformation rule \begin{equation}
\FA\!\left(\gamma\tau\right)=\varrho\!\left(\gamma,\tau\right)\FA\!\left(\tau\right)\label{eq:modtrans}\end{equation}
for all $\gamma\!\in\!\FG$ and $\tau\!\in\!\uhp$; clearly, automorphic
forms are the meromorphic sections of the corresponding vector bundle.
An automorphic form $\FA\!\left(\tau\right)$ is called weakly holomorphic
if it is holomorphic in the upper half-plane $\uhp$, and has only
a finite order pole at $\tau\!=\!\mathsf{i}\infty$, meaning that
its Puisseux series (in terms of the local uniformizer) involves only
finitely many negative powers; if there are no negative powers at
all, then $\FA\!\left(\tau\right)$ is holomorphic, and it is a cusp
form if it vanishes in the limit $\tau\!\rightarrow\!\mathsf{i}\infty$.
Another common appelation for weakly holomorphic forms, spread across
much of the literature, is vector-valued modular form \cite{bib:BG2,bib:ES,bib:KM},
emphasizing their multicomponent nature. We shall denote by $\mro{\varrho}$
the set of weakly holomorphic forms for the automorphy factor $\varrho$;
obviously, these form a linear space over $\mathbb{C}$.

At this point, it could be helpful to review the case of classical
scalar modular forms \cite{bib:Ap,bib:diamond,bib:Knop,bib:koblitz,bib:lang,bib:serre}.
These are automorphic forms for the rank one automorphy factors \begin{equation}
\rho_{2k}\!\left(\gamma,\tau\right)=\left(c\tau+d\right)^{2k}\,\,\label{eq:trivik}\end{equation}
of weight $2k$, defined for any integer $k$. There exist holomorphic
forms only for $k\!>\!1$, classical examples being the Eisenstein
series \cite{bib:Ap,bib:serre} \begin{equation}
E_{2k}\!\left(q\right)=1-\frac{2k}{B_{k}}\sum_{n=1}^{\infty}\sigma_{k-1}\!\left(n\right)q^{n}\,\,,\label{eq:Ekdef}\end{equation}
where \begin{equation}
\sigma_{k}\!\left(n\right)=\sum_{d|n}d^{k}\,\,\label{eq:skdef}\end{equation}
is the $k$-th power sum of the divisors of the integer $n$, and
$B_{k}$ denotes the $k$-th Bernoulli number. Actually, any holomorphic
form may be expressed uniquely as a polynomial in $E_{4}\!\left(\tau\right)$
and $E_{6}\!\left(\tau\right)$. There are no cusp forms for $k\!<\!6$,
and there is a unique one (up to a multiplicative constant) for $k\!=\!6$,
the famous discriminant form \begin{equation}
\Delta\!\left(\tau\right)=\frac{E_{4}\!\left(\tau\right)^{3}-E_{6}\!\left(\tau\right)^{2}}{1728}=q\prod_{n=1}^{\infty}\left(1-q^{n}\right)^{24}\,\,.\label{eq:deltadef}\end{equation}
The expression of $\Delta$ as an infinite product shows that the
discriminant form doesn't vanish on the upper half-plane \cite{bib:Ap,bib:serre},
and this makes it possible to construct weakly holomorphic forms for
arbitrary $k$ as suitable quotients of Eisensteins by powers of $\Delta$.
In particular, the\textcolor{black}{{} so-called Hauptmodul} \cite{bib:Ap,bib:diamond,bib:koblitz}
\begin{equation}
J\!\left(q\right)=\frac{E_{4}\!\left(\tau\right)^{3}}{\Delta\!\left(\tau\right)}-744=q^{\textrm{-}1}+\sum_{n=1}^{\infty}c\!\left(n\right)q^{n}=q^{\textrm{-}1}+196884q+21493760q^{2}+\ldots\,\,,\label{eq:Jexp}\end{equation}
is invariant under $\FG$, i.e. $J\!\left(\gamma\tau\right)\!=\! J\!\left(\tau\right)$
for all $\gamma\!\in\!\FG$, is holomorphic in $\uhp$, and has a
first order pole at the cusp, hence it is a weakly holomorphic form
for $\rho_{0}$, and every element of $\mro{\rho_{0}}$ can be expressed
as a univariate polynomial in $J\!\left(\tau\right)$. Finally, we
note that the Eisenstein series $E_{2}\!\left(\tau\right)$, defined
by Eq.\eqref{eq:Ekdef} for $k\!=\!1$, is equal to the logarithmic
derivative of the discriminant form \begin{equation}
E_{2}\!\left(\tau\right)=\frac{1}{2\pi\mathsf{i}}\dif{\!\left(\ln\Delta\right)}{\tau}\,\,.\label{eq:deltadif}\end{equation}
While obviously holomorphic, $E_{2}\!\left(\tau\right)$ is not a
modular form, since it does not satisfy the transformation rule Eq.\eqref{eq:modtrans}.

A major simplification follows from the observation that the theory
for automorphy factors of non-zero weight may be reduced to the case
of zero weight via the so-called weight shifting trick. Indeed, thanks
to the fact that the discriminant form doesn't vanish on the upper
half-plane, it is meaningful to consider arbitrary fractional powers
of $\Delta$. This allows one to associate to the flat automorphy
factor $\varrho$ of weight $w$ the automorphy factor \begin{equation}
\varrho_{{\scriptscriptstyle 0}}\!\left(\gamma,\tau\right)=\varrho\!\left(\gamma,\tau\right)\left(\frac{\Delta\!\left(\tau\right)}{\Delta\!\left(\gamma\tau\right)}\right)^{\nicefrac{w}{12}}\label{eq:wshift}\end{equation}
of weight $0$. If $\FA\!\in\!\mro{\varrho}$ is a weakly holomorphic
form for $\varrho$, then \begin{equation}
\FA_{{\scriptscriptstyle 0}}\!\left(\tau\right)\!=\!\Delta\!\left(\tau\right)^{\textrm{-}\nicefrac{w}{12}}\FA\!\left(\tau\right)\label{eq:X0}\end{equation}
is a weakly holomorphic form for $\varrho_{{\scriptscriptstyle 0}}$,
providing a one-to-one correspondence between $\mro{\varrho}$ and
$\mro{\varrho_{{\scriptscriptstyle 0}}}$. Consequently, we can restrict
our attention to forms for automorphy factors of weight $0$, without
any loss of generality. What is more, in many important applications
the only automorphy factors that actually show up are of weight $0$.
For example, the character vector of a RCFT, formed by the genus one
characters of the primary fields, is a \wf \cite{bib:BG}, its weight
$0$ automorphy factor being the celebrated modular representation
determined by the fusion rules and conformal weights of the primary
fields via Verlinde's theorem \cite{Ver,MS}.

For the above reasons, from now on we shall only consider flat automorphy
factors of weight $0$, i.e. matrix representations $\varrho\!:\!\FG\!\rightarrow\!\gl d{\mathbb{C}}$.
We make the following technical assumptions on $\varrho$ \cite{bib:BG,bib:BG2}:
\begin{enumerate}
\item $\varrho\!\sm{\textrm{-}1}00{\textrm{-}1}$ equals the identity matrix;
\item there exists a real diagonal matrix $\FB$, called the \textit{\emph{exponent
matrix}}\emph{,} such that \begin{equation}
\varrho\!\sm 1101=\exp\!\left(2\pi\mathsf{i}\FB\right)\:.\label{eq:expdef}\end{equation}

\end{enumerate}
Note that $\FB$ is not unique, since only the fractional part of
its diagonal elements are determined by Eq.\eqref{eq:expdef}, but
not the integer parts.

If $\FA\!\left(\tau\right)$ is an automorphic form for $\varrho$,
then \eqref{eq:expdef} implies, taking into account Eq.\eqref{eq:modtrans},
that the map $\exp\!\left(\textrm{-}2\pi\mathsf{i}\FB\tau\right)\FA\!\left(\tau\right)$
is periodic in $\tau$ with period $1$: as a consequence, it may
be expanded into a Fourier series (the \emph{$q$}-expansion of $\FA$)\begin{equation}
q^{-\FB}\FA\!\left(q\right)=\sum_{n\in\mathbb{Z}}\FA\!\left[n\right]q^{n}\:,\label{eq:qexp}\end{equation}
where $q\!=\!\exp\!\left(2\pi\mathsf{i}\tau\right)$ and $\FA\!\left[n\right]\!\in\!\mathbb{C}^{d}$.
Note that $\FA$ is weakly holomorphic precisely when its $q$-expansion
contains only finitely many negative powers of $q$.

For any automorphy factor $\varrho$, multiplication by $J\!\left(\tau\right)$
takes the linear space $\mro{\varrho}$ to itself: in other words,
$\mro{\varrho}$ is a $\FH$-module, which may be shown to be free
of finite rank. Under suitable circumstances (that are always met
in practice), there exists a $d$-by-$d$ matrix $\om{\tau}$, whose
columns freely generate $\mro{\varrho}$, and such that the asymptotic
relation \begin{equation}
\om q\rightarrow q^{\FB-1}\;\quad\mathrm{as}\,\quad q\rightarrow0\:\,\label{eq:xibc}\end{equation}
holds for a suitable exponent matrix $\FB$. In such case we call
$\om{\tau}$ a fundamental matrix for $\varrho$, and the limit\begin{equation}
\mathcal{X}=\lim_{q\rightarrow0}\left(q^{-\FB}\om q-q^{-1}\right)\,\,,\label{eq:xmdef}\end{equation}
whose existence follows from Eq.\eqref{eq:xibc}, the \textit{\textcolor{black}{\emph{corresponding
characteristic matrix}}} \cite{bib:BG2}. As we shall see later, the
numerical matrices $\mathcal{X}$ and $\FB$ determine the fundamental
matrix completely, and provide a useful parametrization of the different
automorphy factors of weight $0$. 

By definition, the fundamental matrix $\om{\tau}$ satisfies the transformation
rule \begin{equation}
\om{\gamma\tau}=\varrho\!\left(\gamma\right)\om{\tau}\,\,,\label{eq:xitrans}\end{equation}
and for each $\FA\!\in\!\mro{\varrho}$ there exists a vector $\prep{\FA}\!\in\!\FH^{d}$
with components that are polynomials in the Hauptmodul $J\!\left(\tau\right)$,
such that \begin{equation}
\FA\!\left(\tau\right)=\om{\tau}\prep{\FA}\,\,.\label{eq:Jrep}\end{equation}
The vector $\prep{\FA}$ is called the polynomial representation of
the \wf  ~$\FA$, and it will play an important role later.

The fundamental matrix $\om{\tau}$ is the most important piece of
data needed to describe \wf s for an automorphy factor of weight
$0$, so the question is how could one determine it. This will be
achieved through the consideration of the invariant differential operators
to be discussed in the next section, which are also of primary interest
in the study of modular differential equations.

\section{Invariant differential operators}

Suppose that $\varrho$ is a flat automorphy factor of rank $d$ and
weight $w$, and that $\FA\!\in\!\mathcal{M\!\left(\varrho\right)}$
is a \wf ~ for $\varrho$. Except in case $w\!=\!0$, the $\tau$
derivative of $\FA\!\left(\tau\right)$ fails to be a \wf , but this
can be cured by the introduction of a suitable correction term. Indeed,
the expression \begin{equation}
\FJ_{w}\FA=\frac{1}{2\pi\mathsf{i}}\dif{\FA}{\tau}-\frac{w}{12}E_{2}\!\left(\tau\right)\FA\!\left(\tau\right)\,\,\label{eq:DDEF}\end{equation}
is easily shown to be a \wf ~ for the automorphy factor\begin{equation}
\varrho^{\prime}\!\left(\gamma,\tau\right)=\varrho\!\left(\gamma,\tau\right)\dif{\!\left(\gamma\tau\right)}{\tau}\label{eq:rohat}\end{equation}
 of weight $w\!+\!2$. Here\begin{equation}
E_{2}\!\left(\tau\right)=\frac{1}{2\pi\mathsf{i}}\dif{\!\left(\ln\Delta\right)}{\tau}=1-24q-72q^{2}-\ldots\label{eq:E2def}\end{equation}
is the logarithmic derivative of the discriminant form $\Delta\!\left(\tau\right)$,
cf. Eq.\eqref{eq:deltadif}.

An important consequence of \eqref{eq:E2def} is the equivariance
relation\begin{equation}
\FJ_{w+12u}\left(\Delta^{u}\FA\right)=\Delta^{u}\FJ_{w}\FA\,\,,\label{eq:Deqvar}\end{equation}
valid for any $u\!\in\!\mathbb{R}$, which allows the use of the weight
shifting trick Eq.\eqref{eq:X0} discussed in the previous section;
if $\FA\!\in\!\mro{\varrho}$, then \begin{equation}
\left(\FJ_{w}\FA\right)_{{\scriptscriptstyle 0}}=\FJ_{0}\FA_{{\scriptscriptstyle 0}}\,\,.\label{eq:DX0}\end{equation}

When defining higher powers of $\FJ_{w}$, one should take into account
that it increases the weight by $2$; as a result, one should use
the recurrence relation \begin{equation}
\FJ_{w}^{n+1}=\FJ_{w+2n}\circ\FJ_{w}^{n}\,\,,\label{eq:Dpowers}\end{equation}
Note that $\FJ_{w}^{n}$ increases the weight by  $2n$. To get a
differential operator of order $n$ that maps $\mathcal{M\!\left(\varrho\right)}$
to itself, one has to multiply $\FJ_{w}^{n}$ by a scalar modular
form of weight $-2n$. A suitable choice is\begin{equation}
\mathfrak{d}_{n}\!\left(\tau\right)=\frac{E_{4}\!\left(\tau\right)^{n_{3}}E_{6}\!\left(\tau\right)^{n_{2}}}{\Delta\!\left(\tau\right)^{n_{\infty}}}\,\,,\label{eq:dnfactdef}\end{equation}
with $n_{k}$ denoting the value of $n$ modulo $k$, and\begin{equation}
n_{\infty}\!=\!\frac{n+2n_{3}+3n_{2}}{6}\,\,.\label{eq:ninfdef}\end{equation}
With the above choice of prefactors, the operators\begin{equation}
\DO_{n}=\mathfrak{d}_{n}\!\left(\tau\right)\FJ_{w}^{n}\,\,\label{eq:nabladef}\end{equation}
are invariant scalar differential operators, which means that they
act on \wf s component-wise, and that they map $\mathcal{M\!\left(\varrho\right)}$
to itself for each automorphy factor $\varrho$. What is more, any
invariant scalar differential operator may be expressed as a linear
combination of the $\DO_{n}$-s, with coefficients that are polynomials
in the Hauptmodul%
\footnote{This follows from the observation that, for a positive integer $n$,
any weakly holomorphic scalar form of weight $\textrm{-}2n$ is the
product of $\mathfrak{d}_{n}\!\left(\tau\right)$ with a form of weight
$0$, i.e. $\mro{\rho_{\,\textrm{-}2n}}\!=\!\mathfrak{d}_{n}\!\left(\tau\right)\mro{\rho_{0}}$.
In other words, $\mathfrak{d}_{n}\!\left(\tau\right)$ generates $\mro{\rho_{\,\textrm{-}2n}}$
as a $\FH$-module.%
}. In particular, this is true for the products $\DO_{n}\!\circ\!\DO_{m}$.
Some of the relevant multiplication rules read

\begin{align}
\DO_{1}\!\circ\!\DO_{1} & =\left(J\!-\!984\right)\!\DO_{2}\!-\!{\textstyle \frac{1}{6}}\left(5J\!+\!264\right)\!\DO_{1}\nonumber \\
\DO_{1}\!\circ\!\DO_{2} & =\left(J\!+\!744\right)\!\DO_{3}\!-\!{\textstyle \frac{2}{3}}\left(J\!-\!984\right)\!\DO_{2}\nonumber \\
\DO_{1}\!\circ\!\DO_{3} & =\left(J\!-\!984\right)\!\DO_{4}-{\textstyle \frac{1}{2}}\left(J\!+\!744\right)\!\DO_{3}\nonumber \\
\DO_{1}\!\circ\!\DO_{4} & =\DO_{5}-{\scriptstyle \frac{1}{3}}\left(J\!-\!984\right)\!\DO_{4}\nonumber \\
\DO_{1}\!\circ\!\DO_{5} & =\left(J\!-\!984\right)\left(J\!+\!744\right)\!\DO_{6}-{\textstyle \frac{1}{6}}\left(7J\!-\!1704\right)\!\DO_{5}\nonumber \\
\DO_{2}\!\circ\!\DO_{1} & =\left(J\!+\!744\right)\!\DO_{3}\!-\!{\textstyle \frac{1}{3}}\left(5J\!+\!264\right)\!\DO_{2}\!+\!{\textstyle \frac{5}{6}}\left(J\!+\!744\right)\!\DO_{1}\label{eq:Dnmulttable}\\
\DO_{2}\!\circ\!\DO_{2} & =\left(J\!+\!744\right)\!\DO_{4}\!-\!{\textstyle \frac{4}{3}}\left(J\!+\!744\right)\!\DO_{3}\!-\!{\textstyle \frac{1}{9}}\left(5J\!+\!264\right)\!\DO_{2}\nonumber \\
\DO_{2}\!\circ\!\DO_{3} & =\DO_{5}\!-\!\left(J\!+\!744\right)\!\DO_{4}\!+\!{\textstyle \frac{1}{3}}\left(J\!+\!744\right)\!\DO_{3}\nonumber \\
\DO_{3}\!\circ\!\DO_{1} & =\left(J\!-\!984\right)\!\DO_{4}\!-\!{\textstyle \frac{1}{2}}\left(5J\!+\!264\right)\!\DO_{3}\!+\!{\textstyle \frac{5}{2}}\left(J\!-\!984\right)\!\DO_{2}\!-\!{\textstyle \frac{5}{36}}\left(7J\!-\!1704\right)\!\DO_{1}\nonumber \\
\DO_{3}\!\circ\!\DO_{2} & =\DO_{5}\!-\!2\!\left(J\!-\!984\right)\!\DO_{4}\!+\!{\textstyle \frac{1}{3}}\left(5J\!+\!264\right)\!\DO_{3}\!-\!{\textstyle \frac{5}{9}}\left(J\!-\!984\right)\!\DO_{2}\nonumber \\
\DO_{3}\!\circ\!\DO_{3} & =\left(J\!-\!984\right)\!\DO_{6}\!-\!{\textstyle \frac{3}{2}}\DO_{5}\!+\!\left(J\!-\!984\right)\!\DO_{4}\!-\!{\textstyle \frac{1}{18}}\left(5J\!+\!264\right)\!\DO_{3}\,\,.\nonumber \end{align}

Of special importance are the periodicity formula\begin{equation}
\DO_{n+6}\!=\!\DO_{n}\!\circ\!\DO_{6}\,\,,\label{eq:nabla6}\end{equation}
which follows from $\FJ_{{\scriptscriptstyle 12}}\Delta\!=\!0$, and
the recursion formula \begin{equation}
\DO_{1}\!\circ\!\DO_{n}=\mathfrak{a}_{n}\DO_{n+1}-\mathfrak{b}_{n}\DO_{n}\,\,,\label{eq:recurs1}\end{equation}
where $\mathfrak{a}_{n}$ and $\mathfrak{b}_{n}$ denote weakly holomorphic
scalar modular forms of weight $0$ that can be expressed as the following
univariate polynomials in the Hauptmodul:\begin{equation}
\begin{split}\mathfrak{a}_{n} & =\left(J\!-\!984\right)^{n_{2}}\left(J\!+\!744\right)^{\frac{n_{3}\left(n_{3}-1\right)}{2}}\\
\mathfrak{b}_{n} & ={\textstyle \frac{n_{2}}{2}}\left(J\!+\!744\right)+{\textstyle \frac{n_{3}}{3}}\left(J\!-\!984\right)\,\,\end{split}
\label{eq:anbndef}\end{equation}
Note that, in complete accord with Eq.\eqref{eq:nabla6}, the coefficients
$\mathfrak{a}_{n}$ and $\mathfrak{b}_{n}$ depend only on the value
of $n$ modulo $6$.

Since the operators $\DO_{n}$, together with the multiplication-by-$J$
operator \begin{equation}
\begin{split}\FC\!:\!\mathcal{M\!\left(\varrho\right)} & \!\rightarrow\!\mathcal{M\!\left(\varrho\right)}\\
\FA\!\left(\tau\right) & \!\mapsto\! J\!\left(\tau\right)\FA\!\left(\tau\right)\end{split}
\,\,\label{eq:bigJ}\end{equation}
all map the space $\mathcal{M\!\left(\varrho\right)}$ of weakly holomorphic
forms to itself, $\mathcal{M\!\left(\varrho\right)}$ is a module
for the noncommutative ring $\dr\!=\!\mathbb{C}\!\left[\mathbf{J},\DO_{1},\DO_{2},\ldots\right]$.
This module is necessarily of finite rank, since it is already of
finite rank as a $\mathbb{C}\!\left[\mathbf{J}\right]$-module. An
important result is that the ring $\dr$ is generated by the operators
$\mathbf{J},\DO_{1},\DO_{2}$ and $\DO_{3}$, as a consequence of
the relations Eqs.\eqref{eq:nabla6} and \eqref{eq:Dnmulttable},
which allow to express any operator $\DO_{n}$ with $n\!>\!3$ in
terms of $\mathbf{J},\DO_{1},\DO_{2},\DO_{3}$, e.g. \begin{equation}
\DO_{4}={\textstyle \frac{1}{1728}}\left(\DO_{2}\!\circ\!\DO_{2}-\DO_{1}\!\circ\!\DO_{3}+{\textstyle \frac{5}{6}}\!\left(\mathbf{J}\!+\!744\right)\!\circ\!\DO_{3}+{\textstyle \frac{1}{9}}\!\left(5\mathbf{J}\!+\!264\right)\!\circ\!\DO_{2}\right)\,\,.\label{eq:D4expr}\end{equation}

It is clear from the multiplication rules Eq.\eqref{eq:Dnmulttable}
and the commutation relations \begin{align}
\left[\DO_{1},\FC\right]\!=\! & -\left(\FC\!-\!984\right)\!\circ\!\left(\FC\!+\!744\right)\nonumber \\
\left[\DO_{2},\FC\right]\!=\! & -2\!\left(\FC\!+\!744\right)\!\circ\!\DO_{1}\!+\!{\textstyle \frac{1}{6}}\left(\FC\!+\!744\right)\!\circ\!\left(7\FC\!-\!1704\right)\label{eq:Dcom}\\
\left[\DO_{3},\FC\right]\!=\! & -3\!\left(\FC\!-\!984\right)\!\circ\!\DO_{2}\!+\!{\textstyle \frac{1}{2}}\left(7\FC\!-\!1704\right)\!\circ\!\DO_{1}\!-\!{\textstyle \frac{2}{9}}\left(\FC\!-\!984\right)\!\circ\!\left(7\FC\!+\!3480\right)\,\,,\nonumber \end{align}
that the structure of the ring $\dr$ does not depend on the automorphy
factor: the ring $\dr$ is universal, one and the same for all automorphy
factors $\varrho$. This shows that there is a very close connection
between \wf  s (for arbitrary automorphy factors) and representations
of $\dr$.

Let's now turn to the determination of the fundamental matrix. Consider
an automorphy factor $\varrho$ of weight $0$, with fundamental matrix
$\om{\tau}$, characteristic matrix $\mathcal{X}$ and exponent matrix
$\FB$. Since each operator $\DO_{n}$ maps $\mro{\varrho}$ to itself,
applying any $\DO_{n}$ to the fundamental matrix gives a matrix whose
columns are \wf  s: consequently, for each positive integer $n$
there exists matrices $\mathcal{D}_{n}\!\left(\tau\right)$ for which
\begin{equation}
\DO_{n}\om{\tau}=\om{\tau}\mathcal{D}_{n}\!\left(\tau\right)\,\,,\label{eq:Dndef}\end{equation}
and whose matrix elements are weakly holomorphic scalar modular forms,
hence polynomials in the Hauptmodul $J\!\left(\tau\right)$; these
polynomials may be determined explicitly by comparing the $q$-expansions
of both sides of Eq.\eqref{eq:Dndef}. For example, \begin{align}
\mathcal{D}_{1} & \!\left(\tau\right)=\left(J\!\left(\tau\right)\!-\!\mathcal{X}\right)\left(\FB\!-\!1\right)+\FB\mathcal{X}-240\left(\FB\!-\!1\right)\label{eq:DN1}\\
\mathcal{D}_{2} & \!\left(\tau\right)=(J\!\left(\tau\right)\!-\!\mathcal{X})(\FB\!-\!1)(\FB\!-\!{\textstyle \frac{7}{6}})+(\FB\!-\!{\textstyle \frac{1}{6}})\FB\mathcal{X}+504(\FB\!-\!1)(\FB\!-\!{\textstyle \frac{73}{63}})\label{eq:DN2}\end{align}
and\begin{multline}
\mathcal{D}_{3}\!\left(\tau\right)=\left(J\!\left(\tau\right)\!-\!\mathcal{X}\right)(\FB\!-\!1)(\FB\!-\!{\textstyle \frac{7}{6}})(\FB\!-\!{\textstyle \frac{4}{3}})+(\FB\!-\!{\textstyle \frac{1}{3}})(\FB\!-\!{\textstyle \frac{1}{6}})\FB\mathcal{X}\\
-480\left(\FB\!-\!1\right)(\FB^{2}-{\textstyle \frac{101}{40}}+{\textstyle \frac{71}{45}})\label{eq:DN3}\end{multline}

For any given $n\!>\!0$, the relation Eq.\eqref{eq:Dndef} is nothing
but a linear differential equation of order $n$ that the fundamental
matrix has to satisfy. A necessary and sufficient condition for this
infinite sequence of equations to have a common solution is the relation\begin{equation}
\DO_{1}\mathcal{D}_{n}\!\left(\tau\right)+\mathcal{D}_{1}\!\left(\tau\right)\!\mathcal{D}_{n}\!\left(\tau\right)=\mathfrak{a}_{n}\!\left(\tau\right)\!\mathcal{D}_{n+1}\!\left(\tau\right)-\mathfrak{b}_{n}\!\left(\tau\right)\!\mathcal{D}_{n}\!\left(\tau\right)\,\,,\label{eq:Dnrecurs}\end{equation}
which follows from Eq.\eqref{eq:Dndef} and the recursion relation
Eq.\eqref{eq:recurs1}; actually,  only the $n\!=\!1$ and $n\!=\!2$
cases of Eq.\eqref{eq:Dnrecurs} have to be considered, the remaining
ones are consequences of these two. Taking into account the known
expression of the matrices $\mathcal{D}_{n}\!\left(\tau\right)$ as
polynomials in the Hauptmodul, Eq.\eqref{eq:Dnrecurs} reduces to
intricate algebraic relations between the matrices $\FB$ and $\mathcal{X}$,
whose explicit form can be found in \cite{bib:BG2}. Provided these
are satisfied, one may compute the fundamental matrix by solving
the first order linear differential equation \begin{equation}
\DO_{1}\om{\tau}=\om{\tau}\mathcal{D}_{1}\!\left(\tau\right)\,\,,\label{eq:compat}\end{equation}
with boundary condition Eq.\eqref{eq:xibc}: the solution will automatically
solve the higher order equations as well. This means that $\om{\tau}$
is completely determined by the two numerical matrices $\FB$ and
$\mathcal{X}$. 

Besides $\dr$, another ring of interest is $\hdr$, spanned by linear
combinations of powers of $\FJ$ with coefficients given by holomorphic
scalar modular forms, i.e. suitable polynomials in the Eisenstein
series $E_{4}$ and $E_{6}$. As an algebra over $\mathbb{C}$, the
ring $\hdr$ is generated by the operators $\FJ$, $\mathbf{E}_{4}$
and $\mathbf{E}_{6}$, where $\mathbf{E}_{k}$ denotes multiplication
by the Eisenstein series $E_{k}\!\left(\tau\right)$. The basic commutation
relations connecting these generators read \begin{align}
\left[\mathbf{E}_{4},\mathbf{E}_{6}\right] & =0\nonumber \\
\left[\mathbf{E}_{4},\FJ\right]\, & ={\textstyle \frac{1}{3}}\mathbf{E}_{6}\label{eq:hdcom}\\
\left[\mathbf{E}_{6},\FJ\right]\, & ={\textstyle \frac{1}{2}}\mathbf{E}_{4}^{2}\nonumber \end{align}

Note that the elements of $\hdr$ usually don't preserve the weight.
If an element of $\hdr$ happens to change the weight of all modular
forms by $2n$, then we say that it is homogeneous of grade $n$.
This leads to a decomposition\begin{equation}
\hdr=\bigoplus_{n=0}^{\infty}\hdr_{n}\,\,,\label{eq:hdgrad}\end{equation}
the grade $n$ homogeneous subspace $\hdr_{n}$ consisting of the
operators that can be written as linear combinations\begin{equation}
\sum_{k=0}^{n}g_{n-k}\!\left(\tau\right)\FJ^{k}\,\,,\label{eq:hddef}\end{equation}
whose coefficients $g_{k}\!\left(\tau\right)$ are holomorphic forms
of weight $2k$; in particular, $g_{0}$ is constant, while $g_{1}\!=\!0$.
Each homogeneous subspace $\hdr_{n}$ is finite dimensional, the set
\begin{equation}
\set{\mathbf{E}_{4}^{a}\mathbf{E}_{6}^{b}\FJ^{n-2a-3b}}{0\!\leq\! a,b,n-2a-3b}\label{eq:hdrbasis}\end{equation}
providing a basis of it. It follows that the Hilbert series of $\hdr$
reads \begin{equation}
H_{\hdr}\!\left(z\right)=\frac{1}{\left(1-z\right)\left(1-z^{2}\right)\left(1-z^{3}\right)}\,\,.\label{eq:hilbhdr}\end{equation}

\section{Modular differential equations}

\global\long\def\han#1{\mathsf{Ann}^{\mathtt{hol}}\!\left(#1\right)}

\global\long\def\An#1{\EuScript A_{#1}}

\global\long\def\mo#1{\mathfrak{X}_{#1}}

An interesting question, with direct relevance to physics, is to determine
all those invariant differential operators that annihilate a given
\wf ~\cite{bib:GKMO,bib:Mas,bib:Zh}. In other words, one is interested
in the annihilator\begin{equation}
\an{\FA}\!=\!\left\{ \nabla\!\in\!\dr\,|\,\nabla\FA=0\right\} \label{eq:andef}\end{equation}
of the \wf ~$\FA\!\in\!\mro{\varrho}$; note that $\an{\FA}$ is
a (left) ideal of the ring $\dr$ of invariant differential operators.
It is clear that, as a $\FH$-module, $\an{\FA}$ is the inductive
limit of the increasing sequence $\An 1\!\subset\!\An 2\!\subset\!\cdots$,
where $\An n$ denotes the set of those elements of $\an{\FA}$ whose
order (as a differential operator) does not exceed $n$, i.e. which
can be written as a combination $\sum\limits _{k=0}^{n}\mathfrak{f}_{k}\!\left(J\right)\DO_{k}$
with polynomial coefficients $\mathfrak{f}_{k}\!\in\!\FH$.

The basic idea is the following: since $\FA\!\in\!\mro{\varrho}$
implies $\DO_{n}\FA\!\in\!\mro{\varrho}$ for all $n$, one has \begin{equation}
\DO_{n}\FA=\om{\tau}\dx n\label{eq:dxdef}\end{equation}
for some $\dx n\!\in\!\FH^{d}$, where $\om{\tau}$ is the fundamental
matrix of $\varrho$; note that $\dx 0\!=\!\prep{\FA}$ is the polynomial
representation of $\FA$, cf. Eq.\eqref{eq:Jrep}. Let's now consider
a syzygy $\mathfrak{s}\!=\!\left(\mathfrak{s}_{0},\cdots,\mathfrak{s}_{n}\right)$
between the vectors $\dx 0,\ldots,\dx n$, i.e. a linear relation
\begin{equation}
\sum_{k=0}^{n}\mathfrak{s}_{k}\!\left(J\right)\dx k=0\,\,\label{eq:syz}\end{equation}
with polynomial coefficients $\FI k\!\in\!\FH$. Such syzygys form
the syzygy module $\syz n$, which is finitely generated and free
according to Hilbert's syzygy theorem \cite{bib:Eis}. Multiplying
both sides of Eq.\eqref{eq:syz} by $\om{\tau}$ from the left, and
taking into account the definition Eq.\eqref{eq:dxdef}, one gets
the result that the differential operator \begin{equation}
\DO_{\mathfrak{s}}=\sum_{k=0}^{n}\FI k\DO_{k}\,\,\label{eq:syzopdef}\end{equation}
annihilates $\FA$. This shows that the map $\mathfrak{s}\!\mapsto\!\DO_{\mathfrak{s}}$
is a module isomorphism between $\syz n$ and $\An n$. Note that
this isomorphism implies that each $\An n$ is a finitely generated
free $\FH$-module. One may compute a free generating set of $\syz n$
using standard methods of commutative algebra, and the corresponding
differential operators will provide a free generating set of $\An n$.

This is all that is needed if one is only interested in the modular
differential equations satisfied by $\FA$ up to some given order.
If one is interested instead in the structure of the full annihilator,
then one needs to understand the relation of $\An{n+1}$ to $\An n$,
at least for large enough $n$. Luckily enough, this relation is fairly
simple. Indeed, let's denote by $\mo n$ the submodule of $\FH^{d}$
generated by the vectors $\dx 0,\ldots,\dx n$; clearly, these submodules
form an increasing sequence $\mo 0\!\subset\!\mo 1\!\subset\!\cdots$.
But the module $\FH^{d}$ is Noetherian, hence any increasing sequence
of submodules saturates, in the sense that there exists a positive
integer $N$ (the saturation index) such that $\mo n\!=\!\mo N$ for
all $n\!\geq\! N$, and in particular, $\dx n\!\in\!\mo N$ for $n\!>\! N$.
Since $\mo N$ is generated by the vectors $\dx 0,\ldots,\dx N$,
this means that for $n\!>\! N$ there exist univariate polynomials
$\mathfrak{p}_{k}^{\left(n\right)}\!\in\!\mathbb{C}\!\left[J\right]$
such that \begin{equation}
\dx n=\sum_{k=0}^{N}\mathfrak{p}_{k}^{\left(n\right)}\!\left(J\right)\dx k\,\,.\label{eq:maxrel}\end{equation}
Multiplying (from the left) both sides of this equality by the fundamental
matrix $\om{\tau}$, we get the equality \begin{equation}
\DO_{n}\FA=\sum_{k=0}^{N}\mathfrak{p}_{k}^{\left(n\right)}\!\left(J\right)\DO_{k}\FA\,\,,\label{eq:maxdif}\end{equation}
from which we conclude that the operator \begin{equation}
\mathfrak{D}_{n}=\nabla_{n}-\sum_{k=0}^{N}\mathfrak{p}_{k}^{\left(n\right)}\!\left(J\right)\nabla_{k}\,\,\label{eq:maxop}\end{equation}
belongs to $\An n$. What is more, given an element $\DO\!=\!\sum\limits _{k=0}^{n}\mathfrak{f}_{k}\!\left(J\right)\DO_{k}\!\in\!\An n$,
the combination $\DO\!-\!\mathfrak{f}_{n}\!\left(J\right)\mathfrak{D}_{n}$
is of order less than $n$, hence belongs to $\An{n-1}$; it follows
that for $n\!>\! N$ the module $\An n$ is generated by $\An{n-1}$
and $\mathfrak{D}_{n}$, and the full annihilator $\an{\FA}$ is generated
(as a $\FH$-module) by the sequence $\mathfrak{D}_{N+1},\mathfrak{D}_{N+2},\cdots$
and a generating set of $\An N$.

What could be said about the annihilator as a left ideal of $\dr$?
Using the multiplication rules Eq.\eqref{eq:Dnmulttable}, one may
show that $\An{N+3}$ generates the full annihilator. This means that
every modular differential equation satisfied by $\FA\!\in\!\mro{\varrho}$
is a consequence of one corresponding to an element of $\An{N+3}$.
In this respect, much more is true: every modular differential equation
satisfied by $\FA\!\in\!\mro{\varrho}$ is a consequence of one corresponding
to a generator of $\An{N+1}$, so that, in order to have full control
over modular equations, it is enough to determine a generating set
of the latter module. 

Of course, to be able to apply the above ideas, one needs first to
determine the sequence of the $\dx n$-s. In principle, this computation
involves transcendental operations; but, thanks to Eqs.\eqref{eq:recurs1}
and \eqref{eq:compat}, one has the simple algebraic recursion \begin{equation}
\dx{n+1}=\!\frac{1}{\mathfrak{a}_{n}}\left\{ \mathcal{D}_{1}\!\left(J\right)\!+\!\mathfrak{b}_{n}\!-\!\left(J\!-\!984\right)\left(J\!+\!744\right)\dif{}J\right\} \dx n\,\,.\label{eq:dxrecursion}\end{equation}
Indeed, applying both sides of Eq.\eqref{eq:recurs1} to $\FA$, and
taking into account Eqs.\eqref{eq:dxdef} and \eqref{eq:compat},
one arrives at Eq.\eqref{eq:dxrecursion}. It is straightforward to
compute, starting from $\dx 0\!=\!\prep{\FA}$, the sequence of $\dx n$-s
by using the above recursion. Once this has been done, the whole story
boils down to some more or less elementary algebraic manipulations,
as explained above.

In some important applications it is not the annihilator that one
is really interested in, but rather the holomorphic annihilator \begin{equation}
\han{\FA}=\!\left\{ \nabla\!\in\!\hdr\,|\,\nabla\FA=0\right\} \,\,.\label{eq:handef}\end{equation}
The first thing to note is that $\han{\FA}$ inherits a grading from
that of $\hdr$, each homogeneous subspace $\han{\FA}_{n}\!=\!\han{\FA}\cap\hdr_{n}$
being finite dimensional. An important quantity related to this decomposition
is the Hilbert-Poincaré-series \begin{equation}
H_{\FA}\!\left(z\right)=\sum_{n=0}^{\infty}\dim\left(\han{\FA}_{n}\right)z^{n}\,\,,\label{eq:HPdef}\end{equation}
which characterizes the rate of growth, as a function of the grade
$n$, of the number of independent holomorphic operators annihilating
the form $\FA$. Note that $H_{\FA}\!\left(z\right)$ is always majorized
as a power series by $H_{\hdr}\!\left(z\right)=1\!+\! z\!+\!2z^{2}\!+\!3z^{3}\!+\!\cdots$.

For a given grade $n\!\geq\!0$, an element of $\han{\FA}_{n}$ can
be expressed as a linear combination of the operators $\mathbf{E}_{4}^{a}\mathbf{E}_{6}^{b}\FJ^{n-2a-3b}$
for $0\!\leq\! a,b,n-2a-3b$. The coefficients in this linear combination
satisfy a system of linear equations, whose coefficient matrix may
be determined by considering the action of the operators $\mathbf{E}_{4}^{a}\mathbf{E}_{6}^{b}\FJ^{n-2a-3b}$
on the $q$-expansion of $\FA\!\left(q\right)$, and by solving this
system, one gets a basis of $\han{\FA}_{n}$. While this direct approach
is conceptually simple, its computational complexity grows rapidly
with the grade $n$, making it unsuitable to treat but the simplest
cases. 

A more effective approach is based on the following observation. Any
element of the homogeneous subspace $\hdr_{n}$ of grade $n$ can
be decomposed as \begin{equation}
\sum_{{a,b\geq0\atop 2a+3b\leq n}}C_{n}\!\left(a,b\right)\mathbf{E}_{4}^{a}\mathbf{E}_{6}^{b}\FJ^{n-2a-3b}\,\,,\label{eq:hanelm}\end{equation}
where the coefficients $C_{n}\!\left(a,b\right)$ are complex numbers.
If this sum annihilates the form $\FA$, then so does \begin{equation}
\sum_{{a,b\geq0\atop 2a+3b\leq n}}C_{n}\!\left(a,b\right)\dpref n\!\left(\tau\right)\mathbf{E}_{4}^{a}\mathbf{E}_{6}^{b}\FJ^{n-2a-3b}\,\,.\label{eq:hanelm2}\end{equation}
But the operators $\mathfrak{d}_{n}\!\left(\tau\right)\mathbf{E}_{4}^{a}\mathbf{E}_{6}^{b}\FJ^{n-2a-3b}$
don't change the weight, hence they all belong to $\dr$, and being
differential operators of order $n-2a-3b$, they are proportional
to $\DO_{n-2a-3b}$, the only basis element of $\dr$ of that order,
i.e.\begin{equation}
\mathfrak{d}_{n}\!\left(\tau\right)\mathbf{E}_{4}^{a}\mathbf{E}_{6}^{b}\FJ^{n-2a-3b}=\hdpref{n,a,b}\DO_{n-2a-3b}\,\,,\label{eq:hnabdef}\end{equation}
for some prefactors\begin{equation}
\hdpref{n,a,b}=\frac{\mathfrak{d}_{n}\!\left(\tau\right)E_{4}\!\left(\tau\right)^{a}E_{6}\!\left(\tau\right)^{b}}{\mathfrak{d}_{n-2a-3b}\!\left(\tau\right)}\label{eq:hnabexp}\end{equation}
 which are weight $0$ scalar modular forms, hence univariate polynomials
in the Hauptmodul $J\!\left(\tau\right)$; the precise form of these
polynomials is easy to work out. Comparing Eqs.\eqref{eq:dxdef} and
\eqref{eq:hnabdef}, we get \begin{equation}
\mathfrak{d}_{n}\!\left(\tau\right)\mathbf{E}_{4}^{a}\mathbf{E}_{6}^{b}\FJ^{n-2a-3b}\FA=\om{\tau}\hdpref{n,a,b}\dx{n-2a-3b}\,\,,\label{eq:dxnab}\end{equation}
hence the element Eq.\eqref{eq:hanelm} belongs to $\han{\FA}_{n}$
if, and only if the following linear relation holds: \begin{equation}
\sum_{{a,b\geq0\atop 2a+3b\leq n}}C_{n}\!\left(a,b\right)\hdpref{n,a,b}\dx{n-2a-3b}=0\,\,.\label{eq:haneq}\end{equation}
The sequence $\dx 0,\dx 1,\ldots$ can be determined using the recursion
relation Eq.\eqref{eq:dxrecursion}, and the polynomials $\hdpref{n,a,b}\!\in\!\FH$
are known, so Eq.\eqref{eq:haneq} is a linear system for the numerical
coefficients $C_{n}\!\left(a,b\right)$: solving this system, one
gets a basis of $\han{\FA}_{n}$. While conceptually a bit more involved,
this method is much more effective than the direct approach based
on the consideration of $q$-expansions.

\section{A worked-out example: the Ising model}

The Ising model \cite{DiFMS} is the Virasoro minimal model of central
charge $c\!=\!\nicefrac{1}{2}$. Its character vector is known to
be \global\long\def\w#1{\mathfrak{f}_{#1}}
 \begin{equation}
\FA=\frac{1}{2}\left(\begin{array}{c}
\w{}+\w 1\\
\w{}-\w 1\\
\sqrt{2}\w 2\end{array}\right)\,\,,\label{eq:isingchar}\end{equation}
where \begin{align}
\mathfrak{f}\!\left(\tau\right) & =q^{-1/48}\prod_{n=0}^{\infty}\left(1+q^{n+\frac{1}{2}}\right)\,\,,\nonumber \\
\mathfrak{f}_{1}\!\left(\tau\right) & =q^{-1/48}\prod_{n=0}^{\infty}\left(1-q^{n+\frac{1}{2}}\right)\,\,,\label{eq:weberdef}\\
\mathfrak{f}_{2}\!\left(\tau\right) & =\sqrt{2}q^{1/24}\prod_{n=1}^{\infty}\left(1+q^{n}\right)\,\,\nonumber \end{align}
are the classical Weber functions . These satisfy the identities \begin{align}
\mathfrak{f}_{1}^{8}+\mathfrak{f}_{2}^{8} & =\mathfrak{f}^{8}\,\,,\label{eq:weberrel1}\\
\w{}\w 1\w 2 & =\sqrt{2}\,\,,\label{eq:weberrel2}\end{align}
and are related to the Hauptmodul trough\begin{equation}
J+744=\frac{\left(\mathfrak{f}^{24}-16\right)^{3}}{\mathfrak{f}^{24}}=\frac{\left(\mathfrak{f}_{1}^{24}+16\right)^{3}}{\mathfrak{f}_{1}^{24}}=\frac{\left(\mathfrak{f}_{2}^{24}+16\right)^{3}}{\mathfrak{f}_{2}^{24}}\,\,.\label{eq:weberJ}\end{equation}

The character vector Eq.\eqref{eq:isingchar} is a modular form for
the weight $0$ automorphy factor characterized by \begin{align*}
\varrho\!\sm 0{\textrm{-}1}10\! & =\!\frac{1}{2}\!\left(\!\begin{array}{rrr}
1 & 1 & \sqrt{2}\\
1 & 1 & \textrm{-}\sqrt{2}\\
\sqrt{2} & \textrm{-}\sqrt{2} & 0\end{array}\!\right)\\
\varrho\!\sm 0{\textrm{-}1}1{\textrm{-}1}\! & =\!\frac{\zeta}{2}\!\left(\!\begin{array}{rrc}
1 & \textrm{-}1 & \,\,\sqrt{2}\zeta^{3}\\
1 & \textrm{-}1 & \textrm{-}\sqrt{2}\zeta^{3}\\
\sqrt{2} & \sqrt{2} & 0\end{array}\!\right)\,\,,\end{align*}
where $\zeta\!=\!\exp\!\left(\frac{2\pi\mathsf{i}}{48}\right)$. A
suitable exponent matrix reads \begin{equation}
\FB=\frac{1}{48}\left(\begin{array}{ccc}
47\\
 & 23\\
 &  & 2\end{array}\right)\,\,.\label{eq:isingL}\end{equation}

The fundamental matrix\[
\!\left(\!\begin{array}{ccc}
\dfrac{\w{}\!+\!\w 1}{2} & \dfrac{\w{}^{25}\!-\!\w 1^{25}\!-\!25\w{}\!-\!25\w 1}{2} & 8\!\left(\w{}^{17}\w 1^{8}\!-\!\w{}^{24}\w 1\!-\!16\w{}\right)\!+\!\dfrac{\w 2^{7}}{\sqrt{2}}\left(\w{}^{39}\!-\!\w 1^{39}\!-\!16\w{}^{15}\!-\!32\w 1^{15}\right)\\
\\\dfrac{\w{}\!-\!\w 1}{2} & \dfrac{\w{}^{25}\!+\!\w 1^{25}\!-\!25\w{}\!+\!25\w 1}{2} & 8\!\left(\w{}^{17}\w 1^{8}\!+\!\w{}^{24}\w 1\!-\!16\w{}\right)\!-\!\dfrac{\w 2^{7}}{\sqrt{2}}\left(\w{}^{39}\!+\!\w 1^{39}\!-\!16\w{}^{15}\!+\!32\w 1^{15}\right)\\
\\\dfrac{\w 2}{\sqrt{2}} & \textrm{-}\left(25\!+\!\w 2^{24}\right)\dfrac{\w 2}{\sqrt{2}} & \w{}^{15}\w 1^{7}\left(\w{}^{24}\!-\!16\right)\!-\!16\w{}^{24}\dfrac{\w 2}{\sqrt{2}}\end{array}\!\right)\]
for this automorphy factor has been determined in \cite{bib:BG},
with corresponding characteristic matrix \begin{equation}
\mathcal{X}=\left(\begin{array}{ccc}
0 & 2325 & 94208\\
1 & 275 & -4096\\
1 & -25 & -23\end{array}\right)\label{eq:isingX}\end{equation}

The character vector of the Ising model clearly equals the first column
of the fundamental matrix, which implies that its polynomial representation
has the simple form\begin{equation}
\prep{\FA}=\left(\begin{array}{c}
1\\
0\\
0\end{array}\right)\,\,.\label{eq:isingpolrep}\end{equation}
Starting from this, it is straightforward to compute the sequence
of $\dx n$-s using the recursion relation Eq.\eqref{eq:dxrecursion},
leading to the result\[
\dx 0\!=\!\left(\begin{array}{c}
1\\
0\\
0\end{array}\right)\!,\;\dx 1\!=\!\frac{1}{48}\!\left(\begin{array}{c}
240\!-\! J\\
24\\
3\end{array}\right)\!,\;\dx 2\!=\!\frac{1}{768}\!\left(\begin{array}{c}
3J\!+\!1448\\
112\\
\textrm{-}7\end{array}\right)\!,\]
\smallskip{}
\[
\dx 3\!=\!\frac{1}{36864}\!\left(\begin{array}{c}
23648\!-\!51J\\
856\\
107\end{array}\right)\!,\;\dx 4\!=\!\frac{1}{1769472}\!\left(\begin{array}{c}
1275J\!+\!288088\\
\textrm{-}2080\\
\textrm{-}2507\end{array}\right)\!,\ldots\]
From this follows that the saturation index is $N\!=\!2$, i.e. $\mo n\!=\!\mo 2$
for $n\!>\!2$, and that there are no syzygys between the generators
of $\mo 2$, i.e. $\syz 2$ (hence $\An 2$ as well) is trivial. By
expressing the higher $\dx n$-s in terms of $\dx 0,\dx 1$ and $\dx 2$,
one gets \begin{align}
\mathfrak{D}_{3}\!= & \,\nabla_{3}-{\textstyle \frac{107}{2304}}\nabla_{1}+{\textstyle \frac{23}{55296}}\!\left(\FC\!-\!984\right)\nonumber \\
\mathfrak{D}_{4}\!= & \,\nabla_{4}-{\textstyle \frac{107}{2304}}\nabla_{2}+{\textstyle \frac{293}{18432}}\nabla_{1}-{\textstyle \frac{23}{110592}}\!\left(\FC\!+\!744\right)\label{eq:isingann}\\
\vdots\,\,\,\,\,\,\,\,\nonumber \end{align}
Since $\An 3$ is generated by $\mathfrak{D}_{3}$\inputencoding{latin2}\foreignlanguage{magyar}{
(because $\An 2$ is trivial), all modular equations satisfied by
$\FA$ are trivial consequences of the single equation $\mathfrak{D}_{3}\FA\!=\!0$.}

\inputencoding{latin9}As to the holomorphic annihilator $\han{\FA}$,
its Hilbert-Poincaré series reads\begin{equation}
H_{\FA}\!\left(z\right)\!=\! z^{3}\!+\! z^{4}\!+\!2z^{5}\!+\!3z^{6}\!+\cdots\!=\!\frac{z^{3}}{\left(1-z\right)\left(1-z^{2}\right)\left(1-z^{3}\right)}\,,\label{eq:isingHP}\end{equation}
showing clearly that $\han{\FA}$ is generated, as an ideal of $\hdr$,
by a single operator of grade $3$, whose expression is (up to a multiplicative
constant) \begin{equation}
\dpref 3\!\left(\tau\right)^{\textrm{-1}}\mathfrak{D}_{3}=\FJ^{3}-{\textstyle \frac{107}{2304}}\mathbf{E}_{4}\FJ+{\textstyle \frac{23}{55296}}\mathbf{E}_{6}\,\,,\label{eq:isinhhgen}\end{equation}
showing that, up to an irrelevant multiplicative factor, the modular
equation $\mathfrak{D}_{3}\FA\!=\!0$ is a holomorphic equation. This
is precisely the one that follows from the null vector relation for
the characters of the Ising model \cite{bib:GK}.

\section{Summary}

Differential equations satisfied by modular forms have been of interest
since the time of Jacobi. The development of String Theory and two
dimensional CFT has led to major advances in the application of the
theory of modular forms to physics, and the important role of the
modular equations satisfied by them has been clear since the early
days. The work of Zhu \cite{bib:Zh} and of Gaberdiel and Keller \cite{bib:GK}
clarified the relation of modular equations and the structure of the
operator algebra,

The present note addressed the question: given a \wf ~ $\FA$ for
some weight zero automorphy factor $\varrho$, determine all modular
differential equations with (weakly) holomorphic coefficients that
are satisfied by $\FA$. As we have seen, there exist effective algorithmic
techniques, based on the general theory of \wf s, that provide full
control over all such modular equations. 

An important possible use is to the existence problem of RCFT: in
some applications (e.g.\cite{Gaberdiel}), one faces the question
whether RCFTs with given properties (fusion rules, modular properties,
torus partition function, etc.) do exist or not. In such cases one
can use the above methods to check whether there exists potential
character vectors that are consistent with the given data, have non-negative
integral $q$-expansion coefficients and do satisfy suitable modular
differential equations. If no such candidate character vectors can
be found, then one can conclude that no RCFT with a consistent operator
algebra exists with the given properties; on the other hand, for each
consistent candidate character vector the holomorphic modular equation
that it satisfies characterize the null vector relations, hence the
representation theory of the would-be operator algebra.


\begin{thebibliography}{28}
\bibitem{bib:Ap} T. M. Apostol, \textit{Modular Functions and Dirichlet
Series in Number Theory} (Springer, 1990).

\bibitem{modeq1}G. Anderson and G.W. Moore, ''Rationality in conformal
fi{}eld theory'', Commun. Math. Phys. \textbf{117} (1988) 441.

\bibitem{bib:BG} P. Bántay and T. Gannon, {}``Conformal characters
and the modular representation'', \textbf{JHEP 0602} (2006) 005.

\bibitem{bib:BG2} P. Bántay and T. Gannon, {}``Vector-valued modular
functions for the modular group and the hypergeometric equation'',
Commun. Number Th. Phys. \textbf{1} (2008) 637--666.

\bibitem{Cardy}J. Cardy, Nucl. Phys. \textbf{B270}, 186 (1986). 

\bibitem{bib:diamond}F. Diamond and J. Shurman, \emph{A first course
in modular forms} (Springer, 2005).

\bibitem{DiFMS}P. Di Francesco, P. Mathieu and D. Senechal, \emph{Conformal
Field Theory} (Springer, 1997).

\bibitem{bib:ES} W. Eholzer and N.-P. Skoruppa, {}``Modular invariance
and uniqueness of conformal characters'', Commun. Math. Phys. \textbf{174}
(1995) 117--136.

\bibitem{bib:EZ} M. Eichler and D. Zagier, \textit{The Theory of
Jacobi Forms}, Prog. Math. 55 (Birkhäuser, Boston, 1985).

\bibitem{bib:Eis}D. Eisenbud, \emph{Commutative algebra with a view
toward algebraic geometry} (Springer, 1995).

\bibitem{Gaberdiel}M.R. Gaberdiel, ''Constraints on extremal self-dual
CFTs'', \textbf{JHEP 0711} (2007) 087.

\bibitem{bib:GKMO} M. R. Gaberdiel, S. Gukov, C. A. Keller, G. W.
Moore, and H. Ooguri, {}``Extremal $N=(2,2)$ 2D conformal field
theories and constraints of modularity'', arXiv: hep-th/0805.4216.

\bibitem{bib:GK} M. R. Gaberdiel and C. A. Keller, {}``Modular diff{}erential
equations and null vectors'', \textbf{JHEP 0809} (2008) 079.

\bibitem{gunning}R.C. Gunning, \emph{Riemann surfaces and generalized
theta functions} (Springer-Verlag,1976).

\bibitem{bib:Knop} M.I. Knopp, \textit{Modular Functions in Analytic
Number Theory} (Markham, Chicago, 1970).

\bibitem{bib:KM1} M. Knopp and G. Mason, {}``Generalized modular
forms'', J. Number Th. \textbf{99} (2003) 1--28.

\bibitem{bib:KM} M. Knopp and G. Mason, {}``Vector-valued modular
forms and Poincaré series'', Illinois J. Math. \textbf{48} (2004)
1345--1366.

\bibitem{bib:koblitz}N. Koblitz,\emph{ Introduction to elliptic curves
and modular forms} (Springer-Verlag, 1993).

\bibitem{Kohnen}Kohnen, W. and Mason, G., ''On generalized modular
forms and their application'', Nagoya J. Math. \textbf{192} (2008),
119-136. 

\bibitem{bib:lang} S. Lang, \emph{Introduction to modular forms}
(Springer-Verlag, 1976).

\bibitem{bib:Mas} G. Mason, {}``Vector-valued modular forms and
linear differential operators'', Intl J. Number Th. \textbf{3} (2007)
377--390.

\bibitem{modeq2}S.D. Mathur, S. Mukhi and A. Sen, Phys. Lett. \textbf{B213}
(1988) 303.

\bibitem{modeq3}S.D. Mathur, S. Mukhi and A. Sen, Nucl. Phys. \textbf{B318}
(1989) 483.

\bibitem{MS}G. Moore and N. Seiberg, Commun. Math. Phys. \textbf{123},
177 (1989). 

\bibitem{bib:serre}J.-P. Serre, \emph{A course in arithmetic }(Springer,
1973).

\bibitem{bib:Sh} G. Shimura, \textit{Introduction to the Arithmetic
Theory of Automorphic Functions}, (Princeton University Press, 1971).

\bibitem{Ver}E. Verlinde, Nucl. Phys. \textbf{B300}, 360 (1988). 

\bibitem{bib:Zh} Y. Zhu, {}``Modular invariance of characters of
vertex operator algebras'', J. Amer. Math. Soc. \textbf{9} (1996)
237--302.
\end{thebibliography}
\end{document}